\documentclass[amsmath, amssymb, twocolumn, floatfix, superscriptaddress, nofootinbib]{revtex4-2}
\usepackage{float}
\usepackage{graphicx}
\usepackage{dcolumn}
\usepackage{bm}
\usepackage{hyperref}%
\usepackage[separate-uncertainty=true]{siunitx}
\usepackage{xcolor}
\usepackage{makecell}
\usepackage{physics}
\usepackage{bbold}
\usepackage{notoccite}
\usepackage[normalem]{ulem}

\newcommand{\id}{\mathbb{1}}

\newcommand{\PRLsep}{\noindent\makebox[\linewidth]{\resizebox{0.625\linewidth}{1pt}{$\bullet$}}\bigskip}

\makeatletter
\newcommand{\globalcolor}[1]{%
  \color{#1}\global\let\default@color\current@color
}
\makeatother

\makeatletter
\def\p@subsection{}

\makeatother

\definecolor{lg}{RGB}{200,200,200}
\definecolor{cyberpink}{HTML}{FE53BB}

\newcommand{\figref}[1]{Fig.~\ref{#1}}

\begin{document}

\title{Demonstration of universal time-reversal for quantum processes}

\author{P. Schiansky\textsuperscript{+}}\email[Corresponding author: ]{peter.schiansky@univie.ac.at}\affiliation{University of Vienna, Faculty of Physics, Boltzmanngasse 5, 1090 Vienna, Austria}
\author{T. Strömberg\textsuperscript{+}}\affiliation{University of Vienna, Faculty of Physics, Boltzmanngasse 5, 1090 Vienna, Austria}
\author{D. Trillo}\affiliation{Institute for Quantum Optics and Quantum Information, Boltzmanngasse 3, 1090 Vienna, Austria}
\author{V. Saggio}\affiliation{University of Vienna, Faculty of Physics, Boltzmanngasse 5, 1090 Vienna, Austria}
\author{B. Dive}\affiliation{Institute for Quantum Optics and Quantum Information, Boltzmanngasse 3, 1090 Vienna, Austria}
\author{M. Navascués}\affiliation{Institute for Quantum Optics and Quantum Information, Boltzmanngasse 3, 1090 Vienna, Austria}
\author{P. Walther}\email[Corresponding author: ]{philip.walther@univie.ac.at}\affiliation{University of Vienna, Faculty of Physics, Boltzmanngasse 5, 1090 Vienna, Austria}

\date{\today}

\begin{abstract}

    Although the laws of classical physics are deterministic, thermodynamics gives rise to an arrow of time through irreversible processes. 
    In quantum mechanics the unitary nature of the time evolution makes it intrinsically reversible, however the question of how to revert an unknown time evolution nevertheless remains.
    Remarkably, there have been several recent demonstrations of protocols for reverting unknown unitaries in scenarios where even the interactions with the target system are unknown. The practical use of these universal rewinding protocols is limited by their probabilistic nature, raising the fundamental question of whether time-reversal could be performed deterministically.
    Here we show that quantum physics indeed allows for deterministic universal time-reversal by exploiting the non-commuting nature of quantum operators, and demonstrate a recursive protocol for two-level quantum systems with an arbitrarily high probability of success.
    Using a photonic platform we demonstrate our protocol, reverting the discrete time evolution of a polarization state with an average state fidelity of over \SI{95}{\percent}.
    Our protocol, requiring no knowledge of the quantum process to be rewound, is optimal in its running time, and brings quantum rewinding into a regime of practical relevance.
\end{abstract}

\maketitle
\def\thefootnote{+}\footnotetext{These authors contributed equally}
\section{Introduction}\label{sec:introduction}
    In the macroscopic world there is an apparent unidirectionality of processes in time, which stands in contrast to the time-reversal symmetric nature of the underlying laws of physics. This tension was first pointed out by Eddington, who coined the term `arrow of time’ to describe the asymmetry~\cite{eddington2019nature}. In classical physics an arrow of time emerges through the second law of thermodynamics, giving rise to processes which cannot be reversed~\cite{callen1998thermodynamics}. Due to the statistical nature of the law, and the determinism of classical physics, the irreversibility is not fundamental. Indeed, for classical wave mechanics it is well known that the time-evolution of a system can be reversed without any knowledge of the dynamics through a technique called phase conjugation~\cite{kuperman1998phase,tompkin1990time}. In the microscopic quantum realm, however, the ability to perform phase conjugation becomes limited by fundamental quantum noise~\cite{gaeta1988quantum}, due to the non-unitary nature of the process. It has therefore remained an open question whether or not the dynamics of quantum systems can be reversed in a universal manner.
    
    \begin{figure}
        \includegraphics[width=1\linewidth]{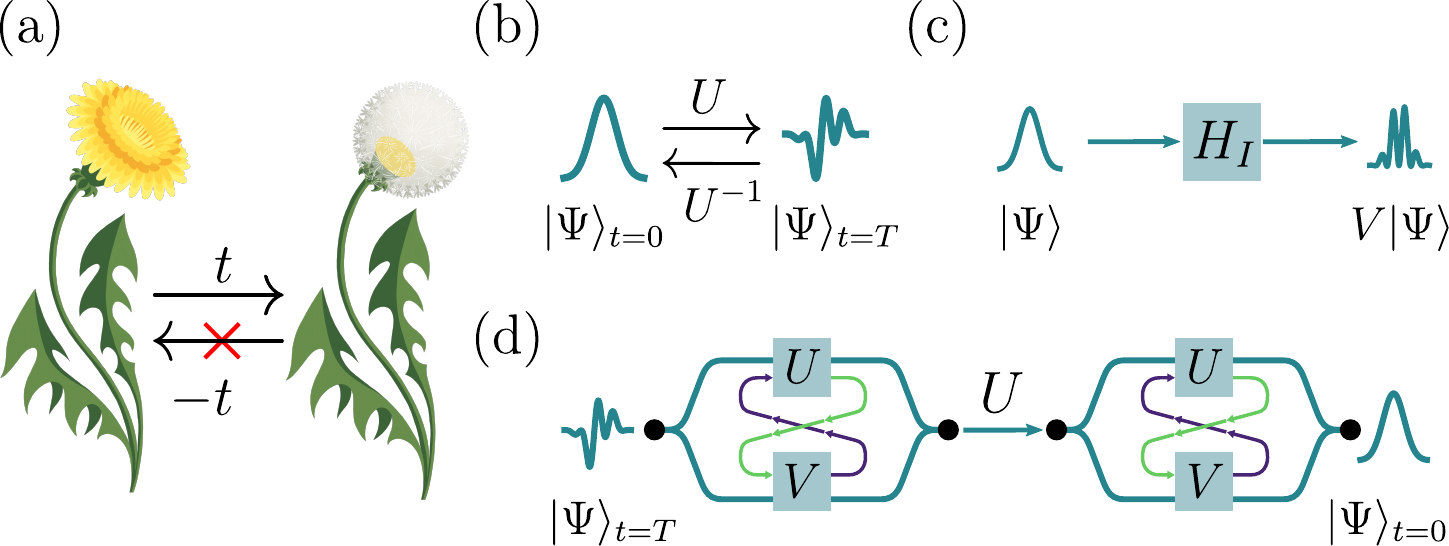}
        \caption{(a)~In the classical world there is an unmistakable directionality to time, illustrated here through the process of ageing; a process which cannot be reversed in practice despite its deterministic nature. In this paper we show that these same limitations do not apply in the quantum realm. (b) The unitarity of quantum mechanics guarantees that an inverse of a given time evolution $U$ always exists, even though it may be unknown. (c) By letting a target quantum system into pass through an interaction region, a perturbed time evolution $V$ can be realised. (d) A quantum SWITCH makes the target system evolve in a superposition of its free evolution $U$ and perturbed evolution $V$. This superposition of time evolutions can be used to `rewind' the system backwards in time, without requiring any knowledge about either $U,V$ or the state $\ket{\Psi}$.}
        \label{fig:figure0}
    \end{figure}
    \begin{figure*}[!t]
        \includegraphics[width=1.0\linewidth]{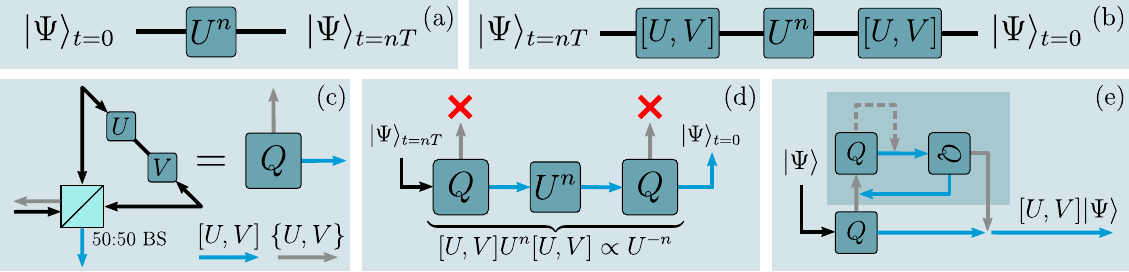}
        \caption{\textbf{Interferometric protocol.} 
        (a) Circuit diagram for the free evolution $U$ of a target quantum from time $t=0$ to $t=T$. (b) Symbolic circuit diagram for the rewinding protocol. The target quantum system is made to propagate backwards in time by way of the identity: $U^{-1}\propto [U,V] U [U,V] $, where $V$ is a perturbed evolution. 
        (c) $Q$ is a quantum SWITCH, pictured here as a Sagnac interferometer, and acts as the basic building block of the interferometric scheme, probabilistically applying the commutator $[U,V]$. 
        (d) A full interferometric implementation of the single-step protocol, which succeeds whenever the photon, in which the target state $\ket{\Psi}$ is encoded, exits both quantum SWITCHes in the commutator port (blue arrow). Detecting a photon in the anticommutator port (grey arrow) heralds a failure of the quantum SWITCH. (e) Adaptive error correction for achieving an arbitrarily high probability of success. This entire diagram replaces a single quantum SWITCH in (d). Instead of detecting the failure mode of the quantum SWITCH the photon is made to re-interfere with itself. Whenever it exits in the bottom right the commutator $[U,V]$ will have been applied (see~\cite{trillo2022})
        The dashed path represents recursive applications of the diagram, through which the success probability can be made arbitrarily high, while the darker shaded area indicates the additional quantum SWITCHes needed.}
        \label{fig:qdiagram}
    \end{figure*}
    Recently there have been several works addressing this question, in which probabilistic protocols for `rewinding' quantum systems were presented~\cite{Navascues2017,Trillo2019} and demonstrated in a lab setting~\cite{Gong2019, Li2019}. These protocols work independently of both the free Hamiltonian guiding the time evolution of the system in question, and the system's interaction with the experimental apparatus. A major drawback of the protocols in~\cite{Trillo2019} is that they suffer from low success probabilities, typically on the order of $10^{-3}$. The scheme in~\cite{Navascues2017}, on the other hand, allows for a form of error correction, whereby the protocol can be repeated when it fails. However, it is not known whether these feed-forward corrections can boost the success probability arbitrarily close to one. Moreover, the protocol cannot rewind a target system in `real-time', instead taking three units of time for every one rewound. 
    
    More traditional methods to rewind a quantum system with an unknown free Hamiltonian, such as the refocusing techniques used in nuclear magnetic resonance \cite{refocusing,Quintino2018}, require the ability to implement controlled operations that are specifically tailored to the target quantum system, and are therefore not universal.
    
    The work of~\cite{sandu} combines both quantum theory and general relativity to devise a `time translator', capable of rewinding or fast-forwarding quantum systems. While this method can time-translate any quantum system, it has two drawbacks: (1) it only works approximately, and under a restriction on the free Hamiltonian of the target; (2) if we demand a reasonable precision, the probability of success of the process becomes astronomically small. 
    
    In this paper we demonstrate a novel universal time-reversal protocol (\figref{fig:figure0}) for which the success probability can be made arbitrarily high, making it, in effect, deterministic. At its heart, the protocol is based on the non-commutativity of quantum operators, a core concept in quantum mechanics. This conceptual simplicity, which translates directly into a straight-forward implementation in the lab based on the recently developed quantum SWITCH~\cite{chiribella2013quantum,procopio2015experimental}, allows us to overcome the limitations of previous proposals.
    
    More specifically, the utilization of quantum SWITCHes allows us to time-translate the unknown internal degree-of-freedom of a target system by setting the target system on a superposition of different trajectories. For some of these trajectories, the free evolution $U$ of the target’s internal degree-of-freedom is perturbed by an unknown but repeatable interaction, which induces an evolution $V$ on the target. This perturbation can be achieved by any physical interaction and thus can be applied to every possible quantum system. We make these trajectories sequentially interfere in such a way that the final state of the target's internal degree-of-freedom is propagated by $U^{-n}$, for some positive $n$, independently of the operators $U, V$. Each quantum SWITCH requires a projection of the target system's path degree-of-freedom to induce the desired superposition of time evolutions. An advantage of our scheme is that even in the event that the projection fails, a simple and repeatable error-correction procedure can be applied, yielding an arbitrarily high success rate, as long as $[U,V]\neq0$. It is also worth emphasizing that the protocol runs in `real-time', meaning that the time it takes to rewind the system is equal to the amount of time to be rewound, aside from a bounded overhead.
    
    We demonstrate the universality of our protocol by running it on a large set of different time evolutions. Our demonstration utilizes a quantum photonics platform with control of path and polarization degrees of freedom of single photons. We generate a discrete time evolution of a single photon by implementing a `polarization Hamiltonian' using a combination of half- and quarter-wave plates. A superposition of time evolutions is achieved via an interferometric quantum SWITCH ~\cite{chiribella2013quantum,procopio2015experimental} in which the propagation direction defines the order of the evolutions $U$ and $V$. Our setup uses two fast optical switches that allow the quantum SWITCH to be accessed several times.

\section{The Protocol}\label{sec: The Protocol}
    In this section we will give a description of how the rewinding protocol works in a photonic setting, the basic steps of which are illustrated in \figref{fig:qdiagram}. An alternative formulation using a scattering scenario is given in the Appendix. A full description, as well as the accompanying proofs, can be found in~\cite{trillo2022}.
    Given an unknown target system $\ket{\Psi}$, whose time evolution is described by $U=e^{-i\Delta TH_0 }$, where $H_0$ is an unknown Hamiltonian, our goal will be to rewind the system: $\ket{\Psi(t=n\Delta T)} \rightarrow \ket{\Psi(t=0)}$ where $n$ is the number of discrete timesteps to be rewound. The basis of our protocol is the following identity:
    \begin{equation}
        [U,V]U^n[U,V]\propto U^{-n}.
        \label{eq:central_equation}
    \end{equation}
    Here $U,V$ are any $2\times2$ matrices, with $U$ being invertible. When the matrix $U$ describes the time evolution of a system, we see that an experimenter able to implement a commutator can reverse the time-evolution, even if $U$ is unknown.
    The basic protocol is thus as follows: apply the commutator between the time evolution operator $U$ and any other $2\times2$ matrix $V$, let the system evolve freely for the amount of time to be rewound, then apply the commutator again. The matrix $V$ represents, in the general setting, a time evolution that is perturbed by any repeatable means, for example by bringing the target on a trajectory that leads it through some interaction region. This perturbed evolution can also remain unknown, however the magnitude of the commutator $[U,V]$ affects the success probability of a single-step attempt to rewind the system. 
    
    In a photonic setting a commutator can be realized using a quantum SWITCH acting on two degrees-of-freedom of a single photon. The control qubit, defining the order of gate operations, is encoded in the photon’s path, while the target qubit is encoded in the polarization. The two possible gate orders, $UV$ and $VU$ are superposed by initializing the control in the superposition state $(\ket{0}_C+\ket{1}_C)/\sqrt{2}$ and then applying a controlled operation between the control and target systems~\cite{procopio2015experimental}:
    
    \begin{figure}
    \includegraphics[width=1.0\linewidth]{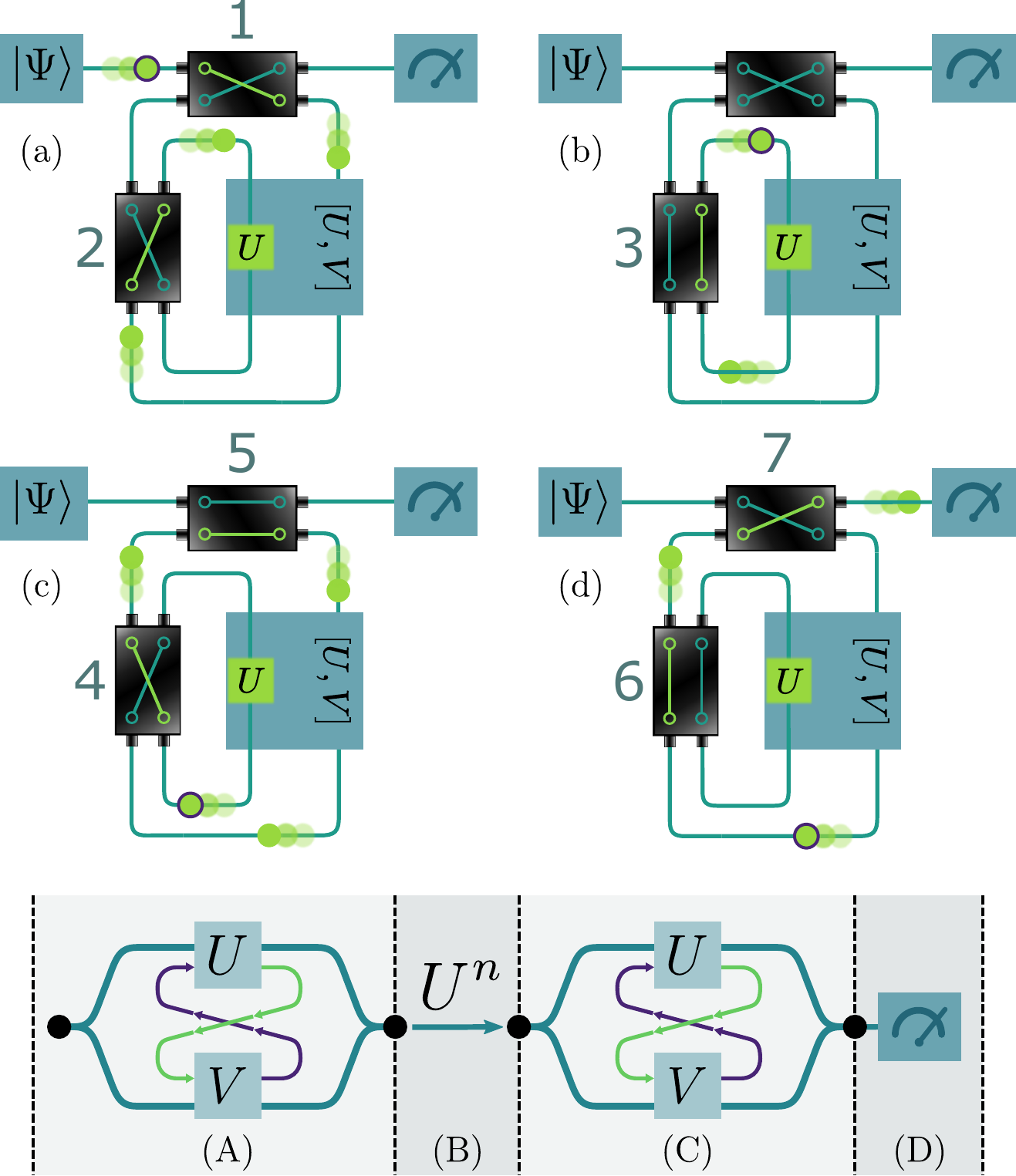}
        \caption{\textbf{Active photon routing.} 
        The use of electro-optical (EO) switches enables active routing of a single photon (green dot) encoding the target quantum state. The settings of the EO-switches determine whether the photon passes through the quantum SWITCH, evolves freely or is sent to a detector.
        The sub-diagrams (a)-(d) indicate the states of the EO-switches at the different steps of the protocol (A)-(D), illustrated at the bottom of the figure.
        The numbers index the order in which the EO-switches are traversed. In each sub-diagram the photon's initial position is indicated by a contour; the subsequent dots represent the photon at a slightly later time, and the green trace shows the photon path through a given switch. (a) The photon passes through the quantum SWITCH for the first time. (b) For $n\geq 2$ the photon is trapped in a loop until the free time-evolution operator $U$ has been applied a total of $n$ times. (c) The photon passes through the quantum SWITCH a second time. (d) The EO-switches direct the photon to a quantum tomography stage.}
        \label{fig:switch_states}
    \end{figure}
    
    \begin{figure*}[!t]
        \includegraphics[width=1.0\linewidth]{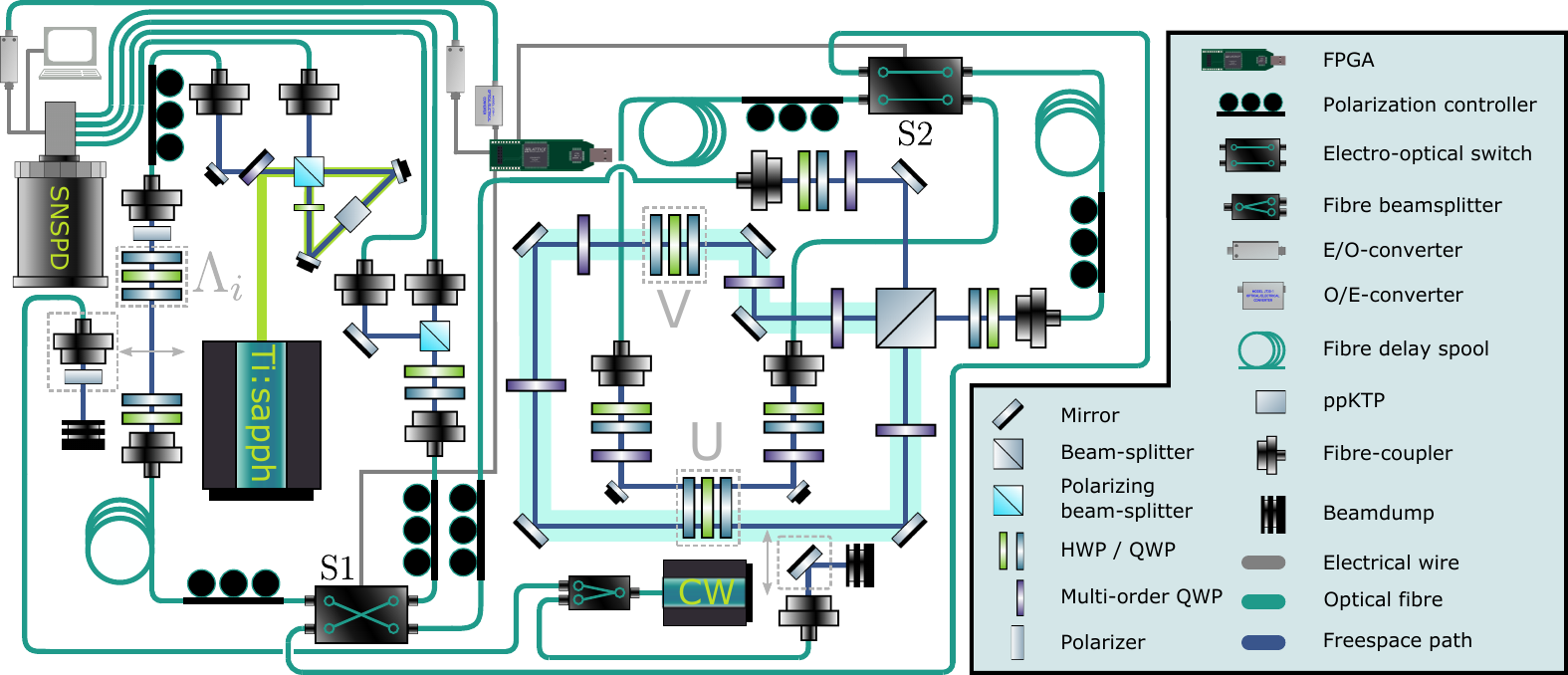}
        \caption{\textbf{Experimental setup.} A pulsed Ti:sapphire laser pumps a spontaneous parametric down-conversion source to generate pairs of single photons in a type-II process using a ppKTP crystal (top left). The signal photon is directed to the high efficiency superconducting nanowire single-photon detectors (SNSPDs), and a successful detection event triggers an FPGA to initiate a pulse sequence for the EO-switches (see \figref{fig:switch_states}). The approximately \SI{400}{\nano\second} rise time of the EO-switches is compensated for by three fiber spools, each around \SI{100}{\meter} long, adding the needed optical delay. 
        The target state, encoded in the idler photon, is initialized using a state-preparation stage after which it is sent to the quantum SWITCH, realised using a free-space Sagnac interferometer (highlighted in blue).
        The unitaries $U$ and $V$ are implemented using a combination of half- and quarter-wave plates. Additional waveplates are used in conjunction with the fiber polarisation controllers to compensate unwanted polarisation rotations induced by the fibers and mirrors. Two additional fiber couplers placed inside the Sagnac allow the photons to propagate through $U$ separately. A tomography stage at the output of EO-switch S1 is used to measure the photons' polarization.
        The CW laser is used during the pre-measurement polarisation compensation procedure.}
        \label{fig:setup}
    \end{figure*}  
    
    \begin{equation}
        \begin{aligned}
            \ket{0}_C  \otimes \ket{\Psi}_T
            \rightarrow
            \frac{\ket{0}_C+\ket{1}_C}{\sqrt{2}} \otimes \ket{\Psi}_T
            \rightarrow\\
            \frac{1}{\sqrt{2}}\big[
            \ket{0}_C \otimes UV\ket{\Psi}_T + 
            \ket{1}_C \otimes VU\ket{\Psi}_T 
            \big]. 
        \end{aligned}
    \end{equation}
    By applying a Hadamard gate to the control qubit, one obtains the following state:
    \begin{equation}
        \label{eq:commutatoridentity}
        \ket{0}_C \otimes \frac{1}{2}\{U,V\}\ket{\Psi}_T +
        \ket{1}_C \otimes \frac{1}{2}[U,V]\ket{\Psi}_T.
    \end{equation}
    A measurement of the control qubit now projects the target state onto either the commutator or the anticommutator, where the latter is denoted by $\{\cdot\}$. If the measurement outcome of the control qubit is $\ket{0}$ the anticommutator is applied by the quantum SWITCH, but the protocol does not necessarily fail. Instead, the following matrix identities can be used to correct the error:
    \begin{align}
        \label{eq:errcorr1}
        \{U,V\}^m [U,V] \{U,V\}^m &\propto [U,V] \\
        \label{eq:errocorr2}[U,V]^2 &\propto \mathbb{1}
    \end{align}
    
    Through recursive application of these identities an anticommutator can always be turned into a commutator. This process can be described using a virtual road map, illustrated in \figref{fig:qdiagram} (e). In~\cite{trillo2022}
    some of us prove that when $U,V$ are unitary and $[U,V]\neq 0$ the protocol always terminates in a finite number of steps. Note that for random $U,V$ the probability of the commutator vanishing is zero. We also point out that Eqs.~\eqref{eq:central_equation},\eqref{eq:errcorr1},\eqref{eq:errocorr2} hold even for non-unitary matrices. 
    Remarkably, the protocol can thus be used to rewind for example a two-level system undergoing a continuous decay governed by a non-Hermitian Hamiltonian.
    
    From \eqref{eq:central_equation} it can be seen that to rewind a free evolution of time $T$ our protocol runs for $T+O(1)$ units of time, which is asymptotically optimal \cite{Trillo2019}. In comparison, the protocol demonstrated in ~\cite{Gong2019, Li2019} takes $3T+O(1)$ units of time for the same task, making our protocol superior not only in terms of success probability, but also in terms of running time.
    
    The above description of the protocol involves placing the target quantum system in a spatial superposition, however we note that the alternative, but equivalent, description of the protocol provided in the Appendix does not require this.
    
\section{Experiment}\label{sec:implementation}
    \begin{figure*}
        \includegraphics[width=1\linewidth]{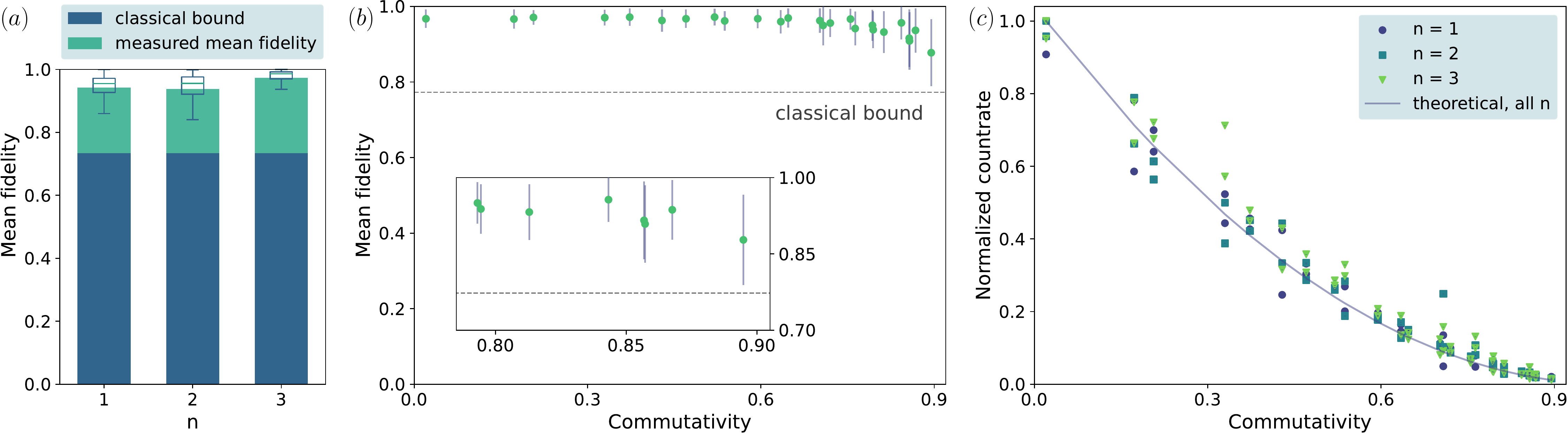}
        \centering
        \caption{\textbf{Experimental results.}
            \textbf{(a) Experimental state fidelities} Each bar shows the measured state fidelity averaged over all 50 pairs of $U,V$, the four different input states and three independent experimental runs, giving a total of 600 different reconstructed density matrices for each value of $n$. The combined measurement time for all $n$ was approximately three weeks. The exact fidelities are $\mathcal{F}_1=({0.94234\pm0.00023}),~ \mathcal{F}_2=({0.93803\pm0.00041}),~  \mathcal{F}_3=({0.97336\pm0.00043})$, with an average of $\mathcal{F}_m=({0.95129\pm0.00021})$.
            The superimposed box plot indicates the median and spread of the fidelities for each $n$. The higher fidelity for $n=3$ can be attributed to higher polarization contrast in the setup (see Appendix A.\ref{sec:methods_expdetails}). The dark blue bars show the highest theoretical fidelity for an experimenter unable to implement superpositions of time evolutions.
            \textbf{(b) Fidelities to $U^{-n}\ket{\Psi_i}$ as a function of the commutativity ($N_c$).} For a given $N_c$ the plotted fidelity is averaged over all runs and input states, with a total of 72 samples per point as several pairs of unitaries commute to the same degree. The error bars show the standard deviation of the fidelities, and not the uncertainty in the estimated mean fidelity, which too small to be visible. At high commutativity the experiment becomes more sensitive to several noise sources, such as detector dark counts, background photons and the leakage of the interferometer due to finite visibility, whereas in the regime of high commutativity the fidelity is limited by constant effects such as finite polarization contrast through the setup. \textbf{(c) Commutativity ($N_c$) versus the normalised total event count rate for all implementations of $V$ and $U$.} Count rates are normalised to the maximal event rate separately for each $n$ (to account for additional losses at higher $n$). The rates are averaged over all 4 input states and all 3 runs. $n=1$ (circles), $n=2$ (rectangles), $n=3$ (triangles). Error bars (Poissonian standard deviation) are too small to be visible. The theoretically ideal behaviour, depicted by the solid line, is given by $N_c^2$ and has a quadratic behaviour due to the commutator being applied twice. The biggest deviation from the overall good agreement to the theory appears in the central region of the curve, where the interferometer has a higher sensitivity to noise.}
        \label{fig:results}
    \end{figure*}
    The rewinding protocol described in the previous section is applied to a qubit state encoded in the polarization degree-of-freedom of a single photon, while the path degree-of-freedom of the same photon is used at two points to encode a second qubit which acts as a control system, thereby enabling the application of a commutator through a controlled unitary inside a quantum SWITCH. The photons are generated using spontaneous parametric down-conversion (SPDC). The SPDC process produces pairs of photons denoted signal and idler, the former of which is sent straight to a detector and is used to herald the presence of the idler photon. Upon such a heralding event a trigger signal is transmitted in optical fiber to an FPGA controlling two active electro-optical (EO) switches to permit the idler photon to pass through parts of the set-up multiple times. 
    The active routing of the photons by the EO-switches is shown in \figref{fig:switch_states}, while a detailed schematic of the setup is displayed in \figref{fig:setup}.
    
    The unitary $\Lambda_{i}$ initializes the idler photon into the polarization state $\ket{\Psi_i}$ chosen from a tomographically complete set, after which an EO-switch (S1) routes the photon into the quantum SWITCH.
    The unitaries $V$ and $U$ inside the quantum SWITCH are implemented using two sets of three wave plates~\cite{simon1990minimal}. Note that there is only one physical realisation of $U$ and $V$, and they could thus in principle remain unknown without compromising the protocol.
    Depending on whether the photon exits in the backwards or forwards propagating port of the interferometer either $[U,V]$ or $\{U,V\}$ is applied. The backpropagating port corresponding to $\{U,V\}$ (\figref{fig:qdiagram} c) is disregarded in our implementation, but could be used to increase the success probability of the protocol.
    Any photon leaving the interferometer in the forward propagating direction passes through a second EO-switch (S2), which traps the photon in a loop allowing it to propagate through $U$ a total of $n$ times. Upon exiting the loop, the photon is directed back to S1 which sends the photon through the quantum SWITCH a second time where $[U,V]$ is applied once more. Finally, the photon is routed to a quantum tomography stage by S1, where its polarisation is measured. These measurements are then used to construct the density matrix $\rho$. In a successful run of the experiment, the state $U^{-n}\ket{\Psi_i}$ is recorded.
    \par
    
\section{Results}\label{sec:results}
    
    To demonstrate that the performance of the protocol is independent of the initial state $\ket{\Psi_i}$, the time-independent evolution $U$, the perturbed evolution $V$, and the number of timesteps $n$, a large set of combinations of these parameters was realised. More specifically, the unitary operators $U$ and $V$ were chosen from the set:
    
    \begin{align*}
        &U_p = e^{-i\arcsin(\alpha)\sigma_z},  & &\alpha = \frac{p}{10},  & & p = 1,\ldots,10, \\
        &V_q = \cos(\theta) \sigma_z + \sin(\theta) \sigma_y,   & &\theta = \frac{q\pi}{11}, & & q=1,\ldots,10.
    \end{align*}
    Depending on the choice of $p$ and $q$ the degree to which the operators $U,V$ commute changes. As a measure of the commutativity we define:
    \begin{equation}
        N_c = 1-\left(\frac{\lvert\lvert[U, V]\rvert\rvert_2 }{2} \right)^{2},
        \label{equ:commutator_norm}
    \end{equation}
    which is normalized to be one when the unitaries are fully commuting and is equal to the probability of applying the commutator in a single step. For our implementation we select the 50 pairs of $U,V$ for which $N_c \leq 0.9$. Choosing a finite set of unitaries generated by fixed Hamiltonians, rather than sampling them randomly, allows us to compare our results to those of a classical experimenter given access to the same resources, but who cannot implement quantum superpositions of time evolutions. The optimal strategy for such a classical experimenter is discussed in the Appendix. While polarization rotations implemented by waveplates alone are in general not invariant under counterpropagation, the specific set above is since it does not contain any $\sigma_x$ terms. We can therefore implement all the unitaries in our set using a Sagnac geometry, without the need for time-reversal symmetry breaking elements. 
    
    To benchmark the fidelity of the protocol we ran it on the four input states $\{\ket{H},\ket{+},\ket{-},\ket{R}\}$, corresponding to horizontally, diagonally, anti-diagonally and right-handed circularly polarized light respectively. This was independently repeated three times for all 50 choices of time evolutions, and for three different sizes of timestep ($n=1,2,3$), yielding a total of 1800 experimental runs with a combined measurement time of more than 500 hours.
    In each experimental run full quantum state tomography was performed on the output states $\rho$, and the fidelity $\bra{\Psi_i}U^{n}\rho U^{-n}\ket{\Psi_i}$ was calculated. The density matrices of the output states were reconstructed using a maximum likelihood fit~\cite{altepeter2005photonic}, and a background contribution originating from the detector dark counts was accounted for using a Monte Carlo simulation, which is how the uncertainties in the fidelities were calculated (see Appendix). The average fidelities for $n=1,2,3$ were   $\mathcal{F}_1=(0.94234\pm0.00023),~
    \mathcal{F}_2=(0.93803\pm0.00041),~ 
    \mathcal{F}_3=(0.97336\pm0.00043)$.
    These fidelities, along with the classical bound, are shown in \figref{fig:results}(a); it can be seen that the quantum protocol clearly outperforms the classical strategy, achieving a high fidelity independent of the length of the time evolution. 
    
    In our implementation the fidelity of the final state is not fully independent of the choice of $U,V$. This is due to the fact that for pairs of unitaries that almost commute, photons are most likely to exit in the anticommutator port of the interferometer, which in turn makes the protocol more sensitive to experimental imperfections such as finite interferometric visibility and detector dark counts. In \figref{fig:results}(b) the relationship between the degree of commutativity $N_c$ and fidelity is illustrated. The mean fidelity stays at high levels over a broad range of $N_c$; only when the degree of commutativity approaches $0.9$ can a small drop in the fidelity be seen.
    
    Since it is expected that the event rate will drop with increasing values of $N_c$, we verify that out setup produces the correct scaling by comparing $N_c$ to our normalized detected photon rate, separately for each $n$.
    The comparison is visualized in \figref{fig:results}(c) where good agreement between relative rate and degree of commutativity can be seen. We attribute the undesired variance in rate to imperfect polarisation compensation inside the Sagnac interferometer, as well as phase shifts originating from slight interferometer misalignment. The largest variance is seen in the neighbourhood around $N_c = 0.5$, where the sensitivity to phase noise is highest, due to the sinusoidal relationship between phase and output intensity in an interferometer.
    
\section{Discussion}\label{sec:discussion}
    In this work we have demonstrated a universal time-rewinding protocol for two-level quantum systems. Unlike previously proposed protocols, ours can reach an arbitrarily high probability of success and is asymptotically optimal in the time required to perform the rewinding, answering the question of whether or not such processes are permitted by the laws of quantum mechanics. Remarkably, the experimenter performing the rewinding does not need any knowledge about the target quantum system, its internal dynamics or even the specifics of the perturbed evolution. The optimality of the protocol is demonstrated in our implementation, where the total elapsed time (equivalent to the number of applications of $U$) grows linearly with the length of time to be rewound, with an optimal proportionality constant of \num{1}. We find that the experimental quantum protocol significantly outperforms the optimal classical strategy in terms of the resulting state fidelity.
    \par
    We emphasize that our results are not restricted to photonic quantum systems, and would be equally applicable to other platforms. However, our photonic implementation offers a particularly simple and robust approach that utilizes a mature technological platform, in particular for implementing the commutator of the time evolutions through a quantum SWITCH. Given the recent progress in integrated quantum photonics ~\cite{wang2020integrated,flamini2018photonic}, we envision that fully monolithic architectures capable of higher fidelity operations will facilitate demonstrations of the active error correction (\figref{fig:qdiagram} (e)) in the near future. Additional follow-up investigations could include non-optical implementations of the protocol as well as extensions to higher dimensions, as described in \cite{Trillo2019}.
\begin{acknowledgments}
    P.S. and T.S. thank Robert Peterson and Lee Rozema for useful discussions. P.W. acknowledges support from the research platform TURIS, the European Commission through EPIQUS (no. 899368) and AppQInfo (no. 956071). from the Austrian Science Fund (FWF) through BeyondC (F7113) and Reseach Group 5 (FG5), from the AFOSR via PhoQuGraph (FA8655-20-1-7030) and QTRUST (FA9550-21- 1-0355), from the John Templeton Foundation via the Quantum Information Structure of Spacetime (QISS) project (ID 61466), and from the Austrian Federal Ministry for Digital and Economic Affairs, the National Foundation for Research, Technology and Development and the Christian Doppler Research Association.
    D.T. is a recipient of a DOC Fellowship of the Austrian Academy of Sciences at the Institute of Quantum Optics and Quantum Information (IQOQI), Vienna. 
\end{acknowledgments}

\section*{Author contributions}
    T.S and P.S. designed and built the experimental setup, carried out the measurements and analyzed the experimental data. D.T. and B.D. provided theoretical support. V.S. assisted with the experimental implementation. M.N. and P.W. supervised the project. All authors contributed to writing the manuscript.

\section*{Competing Interests statement}
    The authors have no competing interests to declare.

\section*{Data availability}
    All data used in the manuscript can be made available upon reasonable request.

\bibliographystyle{unsrt}
\bibliography{bibliography}
    
\onecolumngrid
\vspace{0.5cm}
\PRLsep
\vspace{0.5cm}
\twocolumngrid
\appendix
\section{Supplementary material}\label{sec:appendices}

\subsection{Connection with the probe framework}
    \label{ssec:probes}
        Previous proposals for universal time translation \cite{Navascues2017}, \cite{Trillo2019} are framed in a \emph{scattering scenario}, where the target system, which sits still in a so-called scattering region, is made to sequentially interact with a number of quantum \emph{probes}. These probes are prepared in a controlled lab and then released to the scattering region, where they interact with the target in an uncharacterized but repeatable way.
        As it turns out, the rewinding protocol introduced in the main text can also be realized as a scattering experiment. In the following we explain how this can be achieved. 
        
        Our goal is to implement a quantum SWITCH by means of scattered probes, which the experimenter has full control over. The probe system $P$, together with an ancillary qubit $A$, are initially prepared in the state:
        \begin{equation}
            \ket{\omega}=\frac{1}{\sqrt{2}}\left(\ket{0}_A\ket{\Phi}_P+\ket{1}_A\ket{\Psi}_P\right).
            \label{probe_ancilla_state}
        \end{equation}
        \noindent Here $\ket{\Phi}$ is a state that remains in the lab and does not interact with the target system, while $\ket{\Psi}$ represents a probe state that allows the probe to enter the scattering region and interact with the target.
        
        We assume that the target system $T$, in the absence of probes, evolves via the (unknown) free Hamiltonian $H_0$. Similarly, when the probe remains in the lab, its evolution is governed by the (known) free Hamiltonian $H_P$. Finally, the interaction between the target and a probe in the scattering region is described by the (unknown) Hamiltonian $H_I$. 
        
        Taking the initial state of the target to be $\ket{\psi}$, we let the joint target-probe state $\ket{\omega}_{AP}\ket{\psi}_T$ evolve for time $\Delta T$:
        \begin{equation}
            \frac{1}{\sqrt{2}}\left(\ket{0}_AU\ket{\psi}_T W\ket{\Phi}_P+\ket{1}_A e^{-iH_I\Delta T}\ket{\psi}_T \ket{\Psi}_P\right),
        \end{equation}
        \noindent where $U=e^{-iH_0\Delta T}$ and $W=e^{-iH_P\Delta T}$. Subsequently, we apply the following (probabilistic) operation on the probe system:
        
        \begin{equation}
            \ket{0}\bra{0}_A\otimes\id_P+\ket{1}\bra{1}_A\otimes W\ket{\Phi}\bra{\Xi}_P,
            \label{prob_operation}
        \end{equation}
        \noindent where $\ket{\Xi}$ is any state of the probe with support in the lab. Defining: 
        \begin{equation}
            V=(\bra{\Xi}_P\otimes\id_T)e^{-iH_I\Delta T}(\ket{\Psi}_P\otimes\id_T),
        \end{equation}
        
        \noindent we can now write the joint state as:
        \begin{equation}
            \frac{1}{\sqrt{2}}\left(\ket{0}_AU\ket{\psi}_T+\ket{1}_AV\ket{\psi}_T\right)W\ket{\Phi}_P,
        \end{equation}
        \noindent Taking advantage of the fact that the probe is now in the state $\ket{\Phi}$, i.e. within the lab, we apply the following operation:
        
        \begin{equation}
            \ket{0}\bra{0}_A\otimes (\ket{\Psi}\bra{\Phi}W^{-1})_P+\ket{1}\bra{1}_A(\otimes W^{-1})_P.
        \end{equation}
        yielding the state:
        \begin{equation}
            \frac{1}{\sqrt{2}}\left(\ket{0}_AU\ket{\psi}_T\ket{\Psi}_P+\ket{1}_AV\ket{\psi}_T\ket{\Phi}_P\right).
        \end{equation}
        \noindent Letting the probe and target systems evolve for another $\Delta T$ units of time and applying (\ref{prob_operation}) once more results in the state:
        
        \begin{equation}
            \frac{1}{\sqrt{2}}\left(\ket{0}_AVU\ket{\psi}_T+\ket{1}_AUV\ket{\psi}_T\right)W\ket{\Phi}_P.
        \end{equation}
        \noindent Since the state of the probe factors out, we ignore it from now on. Finally, we measure the ancillary qubit in the basis $\ket{\pm}=\frac{1}{\sqrt{2}}(\ket{0}\pm\ket{1})$. Depending on the measurement result $\pm$, the final state of the target system will be
        
        \begin{equation}
            \frac{1}{2}\left(VU\pm UV\right)\ket{\psi}.
        \end{equation}
        \noindent This is an implementation of the quantum SWITCH gate.
        
    \subsection{Classical strategies}
    In this section we compute the fidelity of a classical rewinding protocol. We consider an experimenter given access to the same resources as a quantum one. Specifically they can choose to either let the system evolve freely for some length of time, or evolve the system using the perturbed time evolution $V$. In contrast to the quantum experimenter, the classical one can only implement these evolutions sequentially, not in a coherent superposition. Thus, the most general classical strategy will have the following form:
        \[
        C(H_0,V,n,t) :=e^{-iH_0t_n}V \cdots e^{-iH_0t_1}Ve^{-iH_0t_0},
        \]
        where $t_j \geq 0$ for all $j$, and
        \[
        n+\sum_{i=0}^n t_i \leq (4+n)\Delta T.
        \]
        This last condition ensures that the classical protocol does not last longer than the one we have used in the main text. Since we are using waveplates as gates, we have to consider a discretized version of time where each gate consumes $\Delta T$ units of time, and thus the most basic rewinding protocol, consisting of implementing $[U,V]U^n[U,V]$, lasts indeed $(4+n)\Delta T$ time units.
        
        Our figure of merit for each $n$ is the average over the fidelities between the final state and the result of rewinding the original state $\ket{\psi}\in {\cal S}$ by an amount $\Delta Tn$. The classical expression is thus
        
        \begin{equation}
            F=\frac{1}{|{\cal P}||{\cal S}|}\sum_{(U,V)\in{\cal P},\psi\in {\cal S}} \abs{\bra{\psi} U^n C(H_0,V,n,t)\ket{\psi}}^2,
        \end{equation}
        \noindent where ${\cal P},{\cal S}$ are, respectively, the set of pairs of operators $(U,V)$ and states $\psi$ considered in the experiment.
        
        We numerically maximize this expression in Mathematica, obtaining $F_c\approx 0.733713$ for $n=1,2,3$. The optimal classical strategy for these choices of states and gates is to let the system evolve unperturbed for $5.91507 - n$ units of time. This particular result is a coincidence, since for other choices the optimal is a non-trivial strategy. 
        
        The numerical optimization over $C$ implies that the classical experimenter posesses knowlege about the set of unitaries. An experimenter restricted to being ignorant about these sets - a constraint we impose on the quantum experimenter - does not have access to this optimal strategy.
    
    \subsection{Experimental details 1}
        Photon pairs centered at approximately \SI{1546}{\nano\meter} are produced in a type-II spontaneous parametric down conversion (SPDC) source based on a periodically poled KTiOPO$_{4}$ crystal in a Sagnac
         configuration~\cite{greganti2018tuning}. The source is pumped by a mode-locked Ti:sapphire laser (Coherent Mira HP) emitting \SI{2}{\pico\second} long pulses at \SI{773.1}{\nano\meter} with a repetition rate of \SI{76}{\mega\hertz}. 
        
        Two electro-optical (EO) switches (Agiltron NanoSpeed) are used throughout the experiment to route the photons in real time, enabling them to pass through the same part of the setup multiple times. Upon the detection of a signal photon, the electrical signal created by the detector is split off, with one copy being amplified to TTL levels using a fast comparator, whereupon it is fed into a waveform preserving electro-optical converter outputting an optical pulse at \SI{1310}{\nano\meter}. The optical signal is sent back through the fiber link to the experimental setup, where it gets re-converted to an electrical signal using an opto-electrical converter, and is then received by the FPGA controlling the EO-switches.
        
        The roughly \SI{400}{\nano\second} rise time of the switches necessitates the use of long fiber delays, which lower the duty cycle of the experiment. These are \SI{519}{\nano\second}/\SI{106}{\meter} between the state preparation $\Lambda_i$ and $S_1$, \SI{533}{\nano\second}/\SI{109}{\meter} between the output of the quantum SWITCH and $S_2$, and \SI{760}{\nano\second}/\SI{155}{\meter} inside of the U-loop (\figref{fig:switch_states}(b)). The \SI{1.5}{\decibel} attenuation per pass through the EO-switches is the main contributor to the overall experimental loss, as the eight passes for $n=3$ add up to \SI{12}{\decibel}. The non-negligible leakage through the switches of around \SI{-13}{\decibel} also contributes to some experimental noise. Additional short fiber delays are used to offset the experimental signal in time to ensure that detection events originating from unused photon pairs are not separated from the real signal by an integer multiple of the pump pulse separation of $\SI{13.2}{\nano\second}$.
        
        The quantum SWITCH is implemented using a bulk Sagnac interferometer to enable long term phase stability. Additionally, the common path geometry ensures that the polarization unitaries are sampled on the same physical spots on the waveplates for both values of the control qubit. The visibility of the interferometer is measured to be in excess of 0.99 for all the four input polarization states used in the experiment. Polarization-dependent phase shifts from the mirrors inside the interferometer are corrected using multi-order QWPs.
    
        Superconducting nanowire single-photon detectors from Photon Spot, housed in a \SI{1}{\kelvin} cryostat, are used for detection. The typical measured detection efficiencies are around \SI{93}{\percent}. An approximately \SI{100}{\meter} long optical fiber link separates the detectors from the experiment. Successful detection events are recorded by a time-tagger with \SI{15.625}{\pico\second} timing resolution.

\subsection{Experimental details 2}\label{sec:methods_expdetails}
    While our SPDC source is able to generate single photons at a rate in excess of \SI{1.5}{\mega\hertz}, many of which cannot be used since a single run of the experiment takes between 2.5 and \SI{4.5}{\micro\second}. Therefore, the FPGA discards all detection events from heralding (signal) photons when a run of the experiment is still in progress. We therefore attenuate the laser pump power until the point where the rate of successful trigger events by the FPGA begins to fall. This also lets us bias the heralding detector at a greater voltage, leading to a higher heralding efficiency.
    
    The Sagnac interferometer constituting the quantum SWITCH is housed in and isolated by three different layers of thinsulate, acrylic and neoprene. This is done to decrease airflow and temperature fluctuations. Gold coated mirrors are used throughout the setup as they exhibit low polarization-dependent loss at our working wavelength (\num{0.034}, \num{0.035} for S, P respectively). Their relatively poor reflectivity of 0.96 adds around \SI{3.5}{\decibel} to the total loss. Similarly, a beam splitter with low polarisation dependent loss and splitting ratio is used for the Sagnac. Multi-order quarter-wave plates are used inside the interferometer to compensate the unwanted polarization-dependent phase shifts caused by the mirrors, with typical polarization contrasts in excess of \SI{40}{\decibel}. After acquiring data for the cases $n=1,2$ we were able to exchange the CW-laser used for this compensation to a model with broader wavelength tuning-range to more closely match our single-photon central wavelength. This lead to superior polarization compensation performance, which in turn explains the increased fidelity $\mathcal{F}_3$ compared to $\mathcal{F}_{1,2}$. While the polarization rotations induced by most components in the setup are not strongly wavelength dependent, the EO-switches are an exception to this, and therefore benefit from a CW-wavelength that more closely matches that of the idler photons.
    \begin{figure*}[t] 
            \centering
            \includegraphics[width=1.0\linewidth]{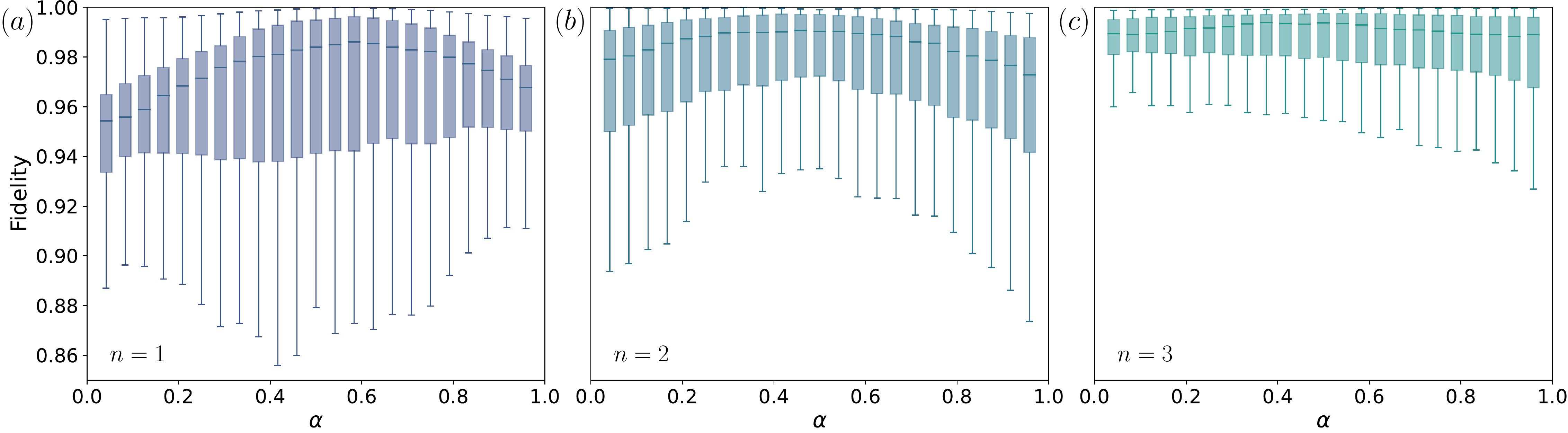}
            \centering
            \caption{\textbf{Fidelities for mixed states.} Density matrices are reconstructed out of convex combinations of data measured for input states $\ket{+}$ and $\ket{-}$. 
            Fidelities to the expected states $U^{-n}\rho U^{-n^{\dagger}}$, with $\rho = \alpha \ketbra{+} + (1-\alpha)\ketbra{-}$, are calculated for varying mixing parameter $\alpha$, all combinations of $(U,V)$, all 3 experimental runs, and all $n$.
            Each box plot shows the median and spread of the state fidelities. \textbf{(a)} $n=1$, \textbf{(b)} $n=2$, \textbf{(c)} $n=3$.}
            \label{fig:mixedstates}
        \end{figure*}
    The polarization unitaries $U_{p}, V_{q}, \Lambda_i$ are implemented with three sets of three waveplates in a QWP-HWP-QWP-configuration, mounted in motorized piezo-electric rotation mounts. While polarization transformations implemented purely with linear retarders will not in general be the same for both propagation directions, our restricted gate set consisting only of linear combinations of $\sigma_y$ and $\sigma_z$ is invariant under change of propagation direction. Since the applications of $U$ outside the quantum SWITCH involve the photons hitting a different spot on the wave-plates, we verify their uniformity by performing quantum process tomography on six randomly generated unitaries. The wave-plates are sampled by four different beams offset horizontally by \SI{2}{\milli\meter} each, and we find that the resulting gate fidelities do not differ by more than \SI{0.1}{\percent}. To verify that the free time-evolution unitaries $U_p$ are faithfully implemented we also perform quantum tomography on them inside the setup and obtain an average fidelity of $\mathcal{F}_U = ({0.9928\pm0.00035})$, averaged over all values of $p$ and the states $\ket{H}$ and $\ket{+}$.
    
    The applicability of the protocol to mixed states is verified by reconstructing density matrices from convex combinations of the acquired data for the pure states $\ket{+}$ and $\ket{-}$. They are compared to expected outcomes for input states of the form $\rho = \alpha \ketbra{+} + (1-\alpha)\ketbra{-}$, for \num{23} values of $ \alpha \in ( 0,1 )$, all 50 combinations of $(p,q)$ and all 3 experimental runs. In \figref{fig:mixedstates}, the fidelities to these expected states are plotted for each value of $n$.
    
\subsection{Signal Processing}\label{sec:methods_noise}
\begin{figure*}[t] 
        \centering
        \includegraphics[width=1.0\linewidth]{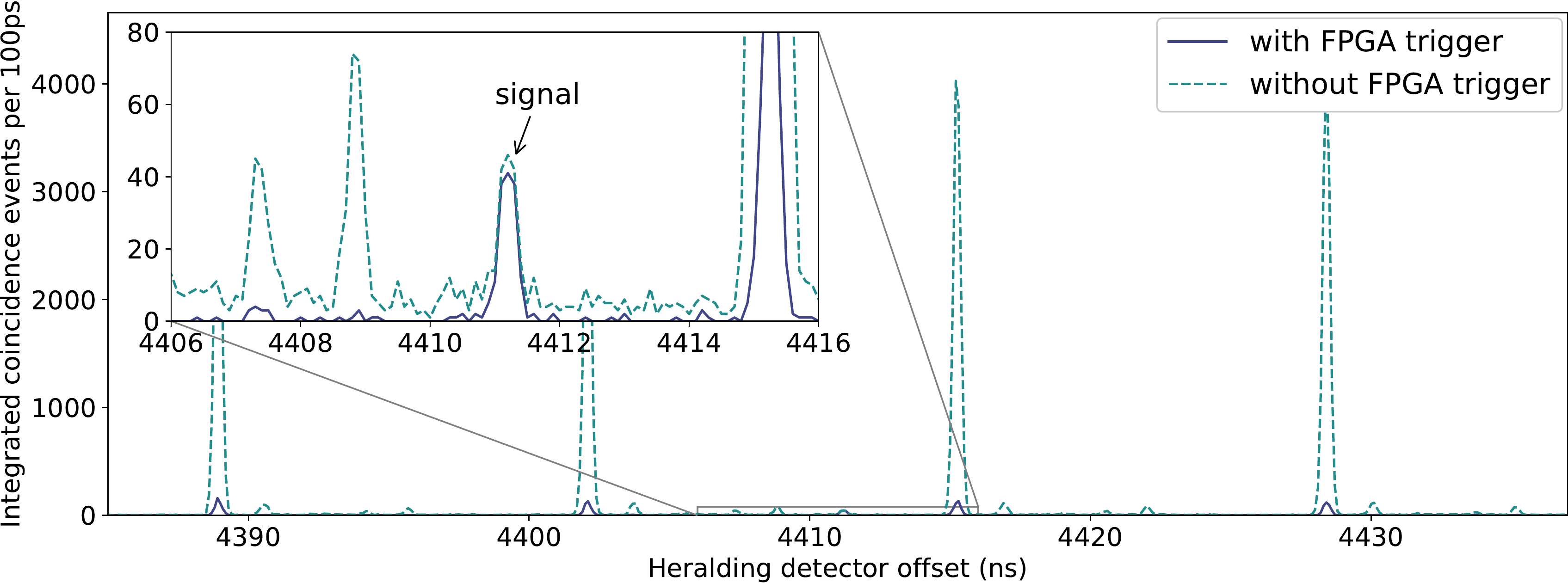}
        \centering
        \caption{\textbf{Influence of FPGA trigger events on noise background.} A comparison between two coincidence histograms with \SI{100}{\pico\second} wide bins taken from the first run of $(p,q,i,n) = (6,2,H,1)$. Background noise originating from higher-order emission events in the SPDC-source as well as detector dark counts can be greatly reduced by filtering out trigger photons that were ignored by FPGA while it was already executing a measurement sequence (blue, solid line). An unprocessed histogram is shown as dashed line. The inset shows a magnified region centered around the signal peak. The suppression of the noise peaks originating from unrelated photon pairs is greater than the ratio of unused to used trigger events. This is because the dead time of the detectors (on the order of \SI{100}{\nano\second}) acts as an additional filter on the heralding photons in the region around our signal, and as a consequence the majority of the events contributing to these peaks in the unfiltered signal comes from heralding photons that were not triggered on.}
        \label{fig:fpga_noise}
    \end{figure*}

    To analyse the data we generate coincidence histograms between the heralding detector and the two detectors connected to the tomography stage.
    As previously stated, due to the long fiber delays the idler photon takes several microseconds to traverse the entire setup. Heralding photons detected within this time window will be ignored by the FPGA, but will still be recorded by the time-tagger. In order to filter out these unused heralding photons in the coincidence analysis, the FPGA outputs a trigger signal whenever it initiates a new pulse sequence. This signal is transmitted back to the time-tagger by the same electro-optic conversion procedure as in the previous section. Only trigger events for which a corresponding signal was received from the FPGA are used in the analysis. Conditioning the photon detection events on the FPGA trigger signal significantly reduces the background noise, as illustrated in \figref{fig:fpga_noise}. 
    
\begin{figure*}[t] 
        \centering
        \includegraphics[width=1.0\linewidth]{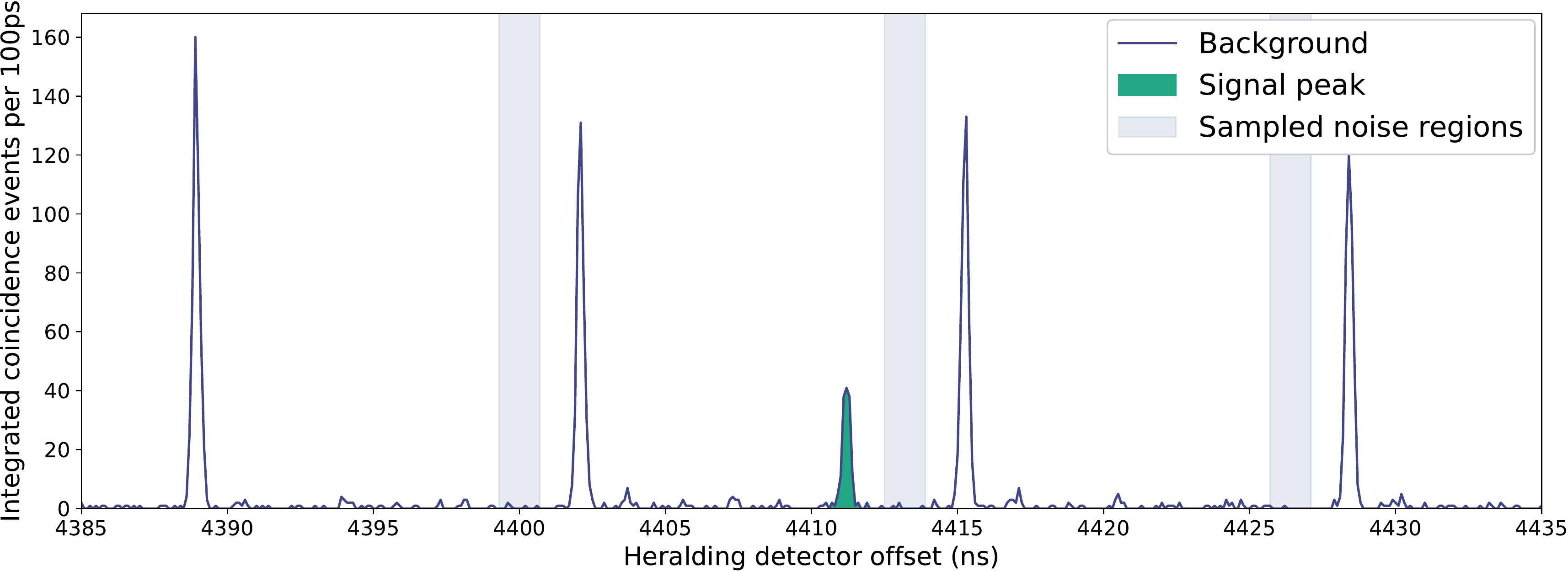}
        \centering
        \caption{\textbf{Background noise sampling.} An example of a coincidence histogram with \SI{100}{\pico\second} wide bins taken from the first run of $(p,q,i,n) = (6,2,H,1)$. It shows the number of coincidence events between the H-port of the tomography stage and the heralding detector, integrated over the measurement time, as a function of a time-offset in the heralding detector. The signal in our experiment is the dark green peak in the center of the graph. Slightly offset from the signal peak one would in an ideal experiment expect zero coincidence events due to the strong time correlation between the signal and idler photons, however due to a small but non-negligible detector dark count rate (on the order of \SI{100}{\hertz}) some coincidence events nevertheless occur. These form a uniform background, which we sample and include as an input to the Monte Carlo simulation that estimates the reconstructed state fidelities. The small side peaks offset from the signal by \SI{13.2}{\nano\second} (which is the reciprocal of the \SI{76}{\mega\hertz} laser repetition rate) are caused by signal photons emitted before or after the pair the FPGAs triggered on. The remaining peaks are caused by coincidences from uncorrelated photon pairs, and this signal is strongly suppressed by conditioning the coincidence counting on the FPGA output signal.}
        \label{fig:histogram}
    \end{figure*}

    Due to the presence of active switches in the setup, in any given run of the experiment there exists multiple possible paths that a photon could have taken from the source to the tomography stage. For example, during the state depicted in \figref{fig:switch_states}~(c) photons can travel directly from the source to the tomography stage. While most such events can be filtered out by virtue of the fact that the difference in arrival time between the signal and idler photons will not match that of the real signal, there are also higher order contributions consisting of signal and idler photons emitted at different times. Since the photons propagating straight to the tomography stage avoid most of the experimental loss, and are not attenuated by the success probability, the rate of these events becomes comparable to the signal even though the intrinsic rate of double-pair emission from the source is significantly lower than the single-pair emission rate. To offset this large noise contribution we add a small fiber delay between S1 and S2. The result is that the signal sits between the major noise peaks, as shown in \figref{fig:histogram}.

    While the SNSPDs have a very low dark count rate, ranging from about 30 to \SI{300}{\hertz}, the high rate of heralding photons nevertheless leads to a small number of accidental coincidence events that form a uniform background in the coincidence landscape. For pairs of implemented unitaries that nearly commute, the resulting low rate of detected signal photons makes the background of accidental coincidences non-negligible. 
    To estimate the impact of this noise in our signal we sample the background in several regions of the coincidence histograms that don't contain any signal. These regions, separated by \SI{13.2}{\nano\second}, are indicated in \figref{fig:histogram}. The mean value of the noise is used as input to a Monte Carlo simulation, from which the mean and standard deviation of the fidelity is obtained. At every step of the simulation a density matrix is constructed using a maximum likelihood fit. The Monte Carlo simulation is allowed to run until the 0.95 confidence interval on the mean fidelity reaches a value below $2\cdot10^{-3}$. In order to further increase the signal to noise, a narrow coincidence window of $0.3-\SI{0.7}{\nano\second}$ is used.

\onecolumngrid
\clearpage

\onecolumngrid
\clearpage

\section{Individual fidelities for all input states, timesteps and pairs of U and V}

    \begin{figure}[ht]
        \centering
        \includegraphics[width=1.0\linewidth]{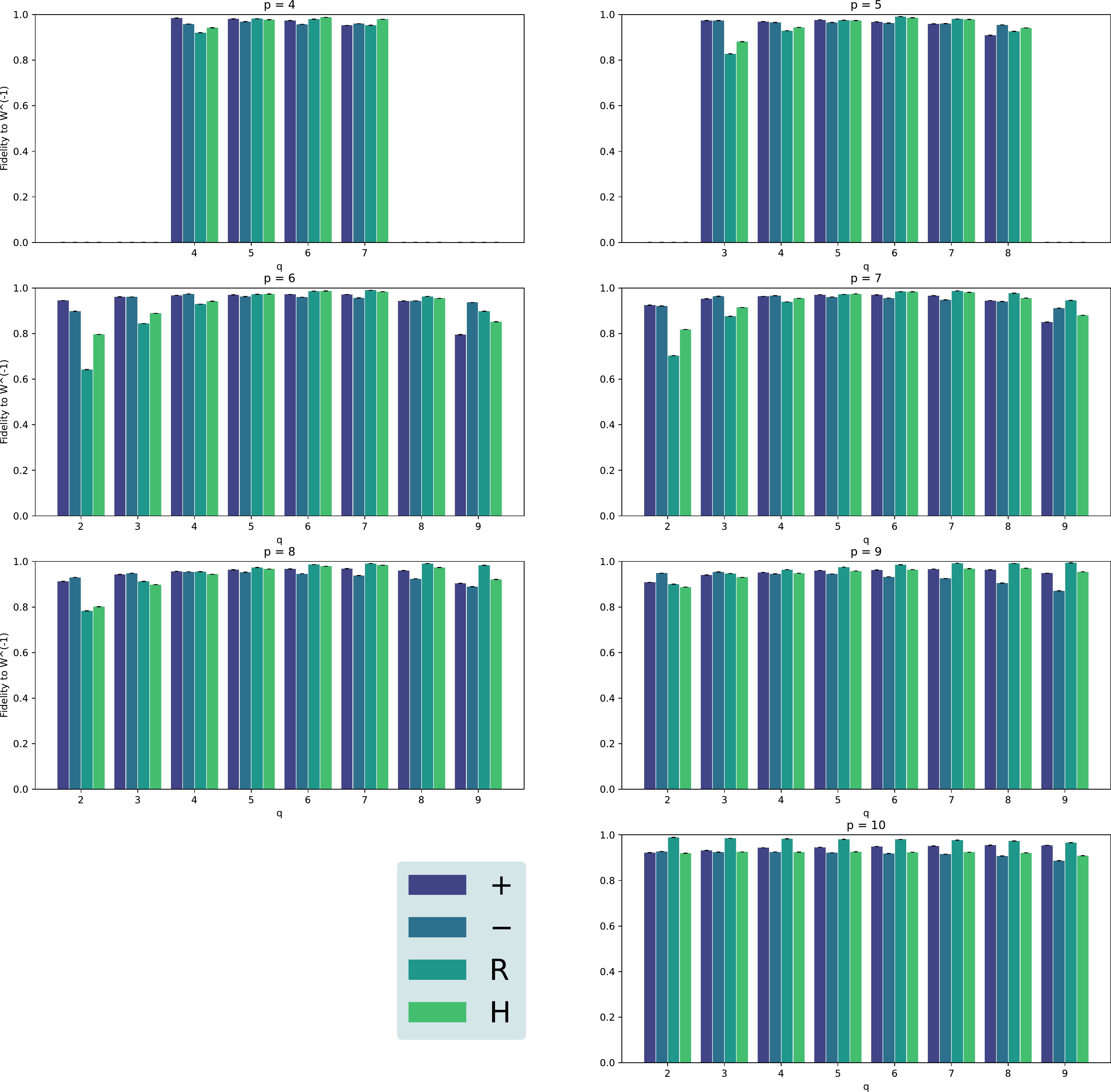}
        \centering
        \caption{\textbf{Fidelities to $U^{-1}\ket{\Psi_i}$.} For all 50 pairs of (U,V) with $N_c \leq 0.9$ and 4 input states each: $+$(violet), $-$(prussian), R(teal), H(green). Averaged over three measurement sets. For each measurement set, the standard deviation of the fidelities calculated by the Monte Carlo simulation is computed. The shown error bars are the root square-sum of these standard deviations.}
        \label{fig:fidelities_s1}
    \end{figure}

    \begin{figure}[ht]
        \vspace{1.35cm}
        \centering
        \includegraphics[width=1.0\linewidth]{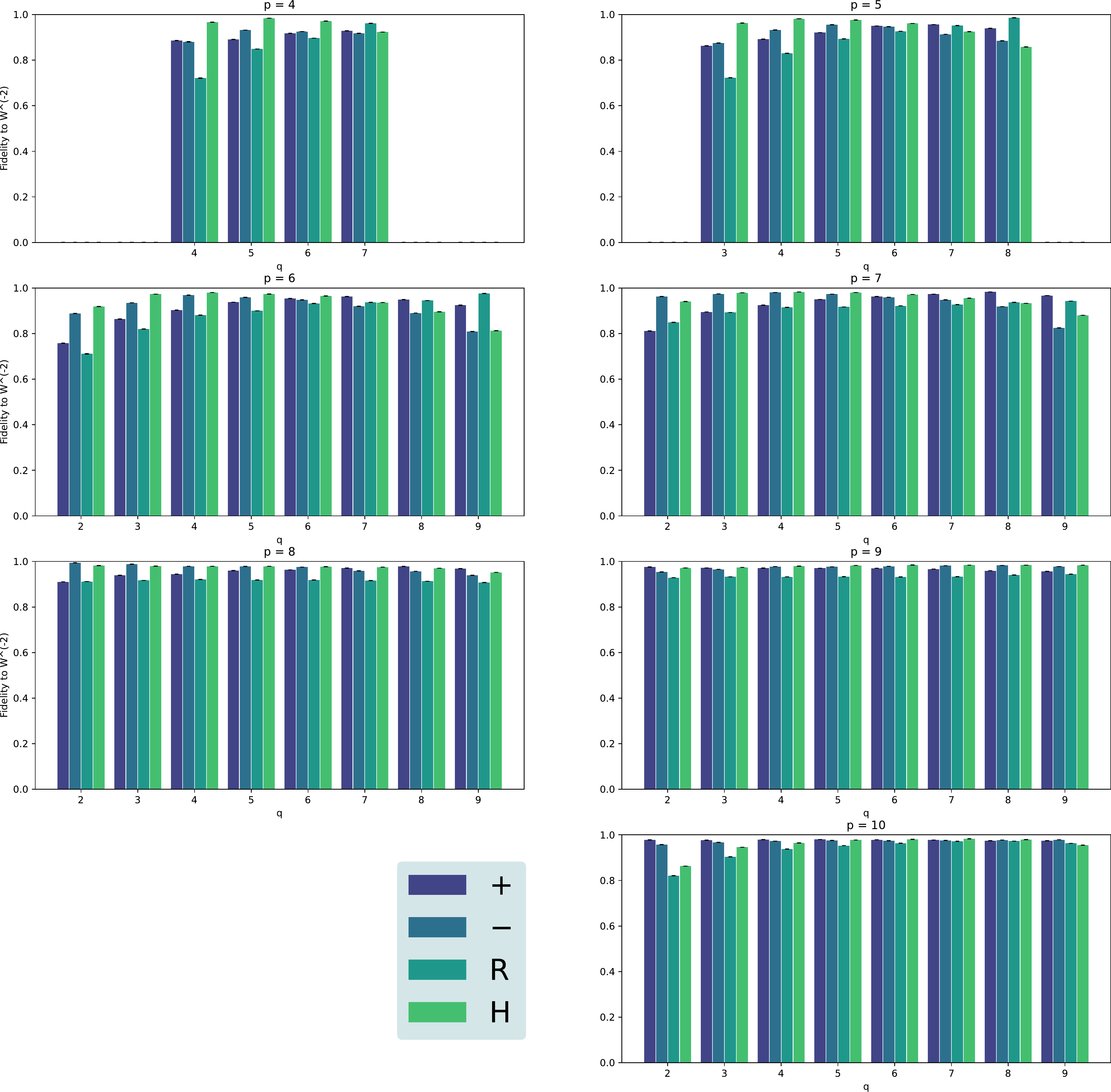}
        \centering
        \caption{\textbf{Fidelities to $U^{-2}\ket{\Psi_i}$.} For all 50 pairs of (U,V) with $N_c \leq 0.9$ and 4 input states each: $+$(violet), $-$(prussian), R(teal), H(green). Averaged over three measurement sets. For each measurement set, the standard deviation of the fidelities calculated by the Monte Carlo simulation is computed. The shown error bars are the root square-sum of these standard deviations.}
        \label{fig:fidelities_s2}
    \end{figure}  

    \begin{figure}[ht] 
        \vspace{1.35cm}
        \centering
        \includegraphics[width=1.0\linewidth]{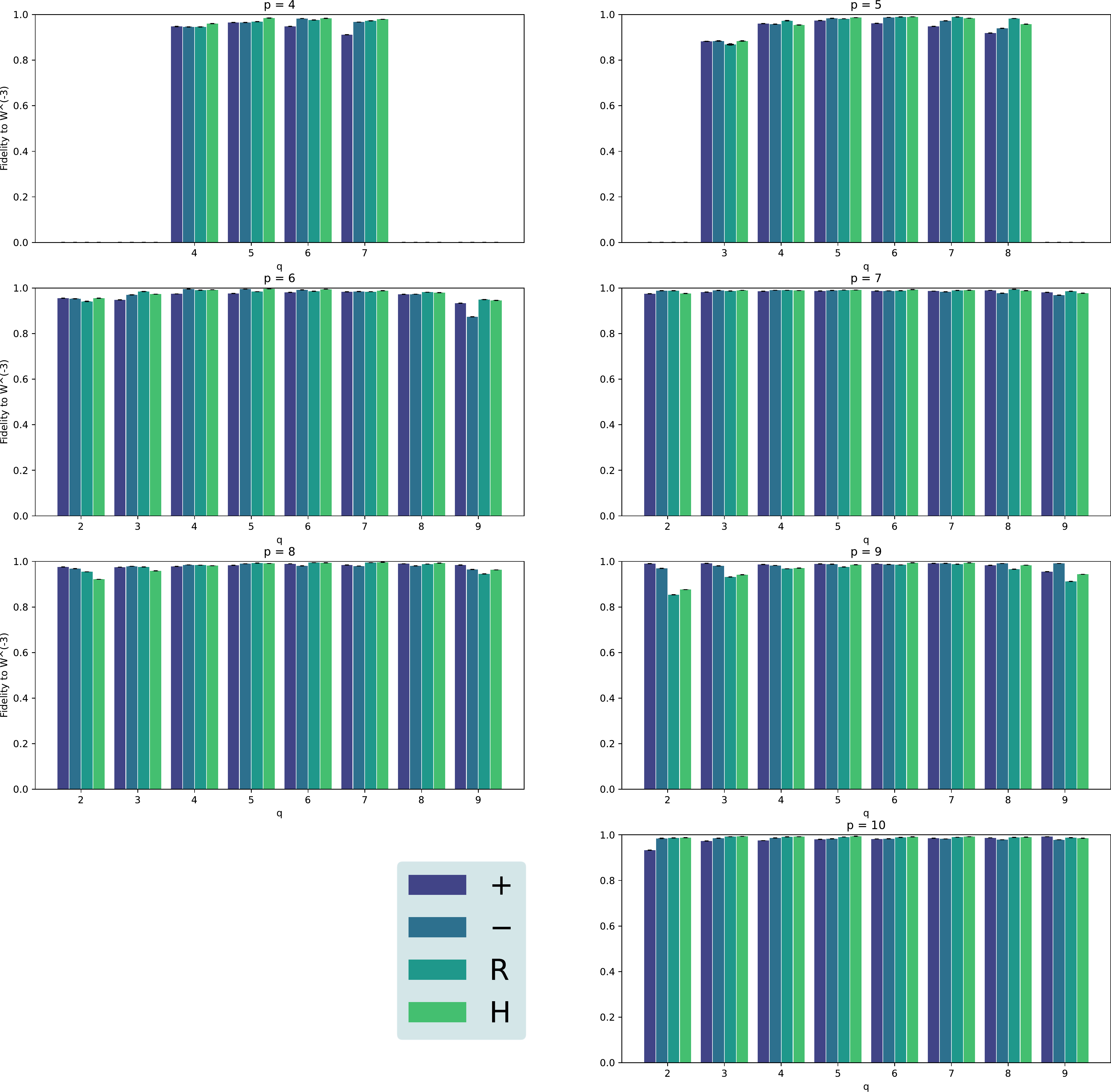}
        \centering
        \caption{\textbf{Fidelities to $U^{-3}\ket{\Psi_i}$.} For all 50 pairs of (U,V) with $N_c \leq 0.9$ and 4 input states each: $+$(violet), $-$(prussian), R(teal), H(green). Averaged over three measurement sets. For each measurement set, the standard deviation of the fidelities calculated by the Monte Carlo simulation is computed. The shown error bars are the root square-sum of these standard deviations.}
        \label{fig:fidelities_s3}
    \end{figure}
    
    \twocolumngrid
    
    \clearpage
    \onecolumngrid
\end{document}